\documentclass[11pt]{article}
\def\be{\begin{equation}}
\def\ee{\end{equation}}
\def\ba{\begin{eqnarray}}
\def\ea{\end{eqnarray}}
\def\la{\langle}
\def\ra{\rangle}
\def\1b{\overline{1}}
\def\a{\alpha}
\def\b{\beta}

\def\h{\hskip 1cm}

\def\lo{\longrightarrow}
\def\A1{A_{-1}}
\usepackage{epsfig}
\begin{document}
\begin{titlepage}
\vspace{4cm}
\begin{center}{\Large \bf Exact ground states for two new spin-1 quantum
chains,\\ new features of matrix product states
}\\
\vspace{2cm}\h S. Alipour\footnote{email:
salipour@physics.sharif.ir},\h V. Karimipour \footnote{email:
vahid@sharif.edu},\h L.
Memarzadeh\footnote{Corresponding author, email: laleh@physics.sharif.edu}, \\
\vspace{3cm} Department of Physics, Sharif University of Technology,
\\P.O. Box 11155-9161, Tehran, Iran

\end{center}
\vskip 3cm
\begin{abstract}
We use the matrix product formalism to find exact ground states of
two new spin-$1$ quantum chains with nearest neighbor interactions.
One of the models, model I, describes a one-parameter family of
quantum chains for which the ground state can be found exactly. In
certain limit of the parameter, the Hamiltonian turns into the
interesting case $H=\sum_i ({\bf S}_i\cdot {\bf S}_{i+1})^2$. The
other model which we label as model II, corresponds to a family of
solvable three-state vertex models on square lattices. The ground
state of this model is highly degenerate and the matrix product
states is a generating state of such degenerate states. The simple
structure of the matrix product state allows us to determine the
properties of degenerate states which are otherwise difficult to
determine.  For both models we find exact expressions for
correlation functions.
\end{abstract}

\end{titlepage}

\section{Introduction}
Quantum information theory and condensed matter physics, study many
body systems on lattices from complementary points of view. While in
condensed matter physics, one starts from a Hamiltonian and seeks to
determine the ground state in some approximate form, in quantum
information theory the emphasis is on the properties of quantum
states, for the quantification of which many tools have been
developed in recent years. The subject of Matrix Product States
(MPS) lies at the borderline of these two disciplines, since in this
formalism, one starts from a quantum many body state with
pre-determined properties, and then constructs a family of local
Hamiltonians for which this state is an exact ground state. In this
way, one may find interesting many-body systems for which the ground
state and all its properties, i.e. correlation functions, can be
calculated in closed analytical form, a very rare situation which is
usually welcomed in condensed matter and statistical physics. \\

The subject of MPS has a long history in condensed matter physics,
the origins of which can be traced back to the work of
Majumdar-Ghosh models \cite{majumdar} which in turn inspired the
construction of a larger family of solvable spin systems by Affleck,
Kennedy, Lieb and Tasaki (AKLT) in \cite{aklt}. The AKLT
construction was further developed in \cite{fcs1,fcs2} under the
name of finitely correlated spin chains or in \cite{zit1,zit2} under
the name of optimal ground states. In its simplest version, which
applies to translational-invariant systems on rings of $N-$sites, a
matrix product state generalizes a product uncorrelated state by
replacing numbers by matrices in the following way
\begin{equation}\label{psigeneral}
    |\psi\ra=\sum_{i_1,i_2,\cdots i_N=1}^d tr(A_{i_1}A_{i_2}\cdots A_{i_N})|i_1,i_2,\cdots i_N\ra,
\end{equation}
where $A_{i}$, $i=1,\cdots d$ are a set of matrices, assigned to the
local states $|1\ra,\cdots |d\ra$ of a site. The normalization of
these states is given by \be\label{norm}\la \psi|\psi\ra=tr(E^N),\ee
 where $ E:=\sum_{i=1}^d A_i^*\otimes A_i $.
The dimensions of these matrices are arbitrary and are constrained
by symmetry considerations and the details of model construction,
i.e. the range of interaction. One can collect all the matrices in a
vector-valued matrix ${\cal A}$ as follows
\begin{equation}\label{calA}
    {\cal A}=\sum_{i=1}^d A_i|i\ra,
\end{equation}
and write the matrix product state (\ref{psigeneral}) as
\begin{equation}\label{psimps}
    |\psi\ra = tr({\cal A}\otimes {\cal A}\cdots\otimes {\cal A}),
\end{equation}
where the trace is taken over the matrix indices and the tensor
product acts on basis vectors, i.e. $$tr({\cal A}\otimes {\cal
B}):=\sum_{i,j} ({A_i})_{\a,\b}({B_j})_{\b,\a}|i,j\ra.$$

The simple structure of the state (\ref{psimps}) allows an exact
calculation of correlation functions. For example one and two-point
functions of local operators are given by
\begin{eqnarray}\label{corrgeneral}
    \frac{\la \psi|O|\psi\ra}{\la \psi|\psi\ra} &=& \frac{tr(E_OE^{N-1})}{tr(E^N)},\cr
    \frac{\la \psi|O_kO_l|\psi\ra}{\la\psi|\psi\ra} &=&
    \frac{tr(E^{k-1}E_OE^{l-k-1}E_OE^{N-l})}{tr(E^N)},
\end{eqnarray}
where \be\label{Eo} E_O=\sum_{i,j} A_i^* \otimes A_j \la i|O|j\ra.
\ee In the thermodynamic limit $N\rightarrow \infty$, the right hand
sides of the above equations simplify even further, since in this
limit, the eigenvalue(s) of $E$ with largest magnitude dominate the
traces.\\

In recent years, this formalism has been used in developing exactly
solvable models in spin chains \cite{zit1, zit2, aks2, aks3,
kolezhukPRB,Rommer, Ian}, spin ladders \cite{zitladder, PolishPaper,
aks1, delgado}, spin systems on two dimensional lattices
\cite{zit2dsquare, zit2dtriangular, peps}, and the study of
entanglement properties of spin systems near the points of quantum
phase transitions \cite{ver1, mpscirac, indian, akm1}. It has also
been used extensively to find the stationary states of many types of
stochastic systems of interacting particles in one dimensional
chains, see for example \cite{derrida, stoch1, karimipour}. \\

A basic question is then whether we can construct general MPS and
its parent Hamiltonians having a set of specific symmetries for
quantum chains of spins. For spin-one systems, the first model was
given by Affleck, Kennedy, Lieb, and Tasaki in \cite{aklt}, (not
within the MPS formalism) which had full rotational symmetry, and
was shown later to correspond to the matrix
\begin{equation}\label{calAaklt}
    {\cal A}_{aklt} = \left(\begin{array}{cc} |0\ra & -\sqrt{2}|1\ra \\ \sqrt{2}|\1b\ra &
    -|0\ra\end{array}\right),
\end{equation}
where $|\1b\ra=|-1\ra$ with the parent Hamiltonian
\begin{equation}\label{Haklt}
    H=\sum_{i} {\bf S}_i\cdot {\bf S}_{i+1} + \frac{1}{3}({\bf
    S}_i\cdot {\bf S}_{i+1})^2.
\end{equation}
Then it was shown \cite{zit1} that if one demands only rotational
symmetry around the $z$ axis in spin space, in addition to parity
and spin-flip symmetries, a more general model can be constructed
which is described by the matrix
\begin{equation}\label{calAzit}
    {\cal A} = \left(\begin{array}{cc} |0\ra & -\sqrt{g}|1\ra \\ \sqrt{g}|\1b\ra &
    \sigma|0\ra\end{array}\right),
\end{equation}
where $g$ is a continuous parameter and $\sigma =\pm 1 $ is a
discrete parameter. \\

At first sight, construction of a matrix product state, finding its
parent Hamiltonian, and calculating the correlation functions, seems
a straightforward procedure. However if one demands symmetry
properties, and more importantly demands that the final Hamiltonian
have a physically interesting interpretation, then the problem will
be quite non-trivial and interesting. Specially if one puts the
formal MPS under scrutiny, one may be able to find many more states
which are not themselves MPS representable, but have been captured
by a single MPS in a nice way, i.e. as their generating state. This
is what we will find for the models constructed in this paper. We
believe that the richness of matrix product formalism has yet to be
unraveled by studying more and more examples. In this paper we try
to construct two other spin-1 matrix product models, which have not
been reported in the literature of matrix product states. The first
model is a one-parameter family which has the interesting property
to interpolate between two limits, namely between the Ising-like
Hamiltonian
\begin{equation}H_1=\sum_{i}(S^2_{z,i}-1)(S^2_{z,i+1}-1),\end{equation}
and
\begin{equation}\label{sss2}H_2=\sum_{i}({\bf S}_i\cdot{\bf S}_{i+1})^2.\end{equation}
The ground state of (\ref{sss2}), as we will see, breaks the
rotational symmetry of the Hamiltonian. This is an example of the
richness of matrix product formalism, that is by searching the space
of solutions, one may come to corners where there are very simple
and physically interesting Hamiltonians whose ground states are
given by MPS. Clearly the symmetry breaking MPS is not the unique
ground state of (\ref{sss2}), however the other ground states, can
be found by applying the symmetry operators of $su(2)$ group to the
MPS.

The other model that we find, labeled as model II, turns out to
correspond to a family of solvable three-state vertex models on
square lattices, \cite{IdzumiRep1, KlumperRep1}. In this model the
degeneracy of the ground states, shows itself in a completely
different way, namely, we find that the matrix product depends on a
continuous parameter, but the parent Hamiltonian does not, i.e.
\begin{equation}\label{}
    H|\psi(g)\ra =E_0|\psi(g)\ra.
\end{equation}
Thus if we expand the matrix product state $|\psi(g)\ra$ in terms of
the parameter $g$, in the form
\begin{equation}\label{}
    |\psi(g)\ra = \sum_{n=0}^N g^n |\psi_n\ra,
\end{equation}
we obtain a large number of states $|\psi_n\ra$, which all have the
same energy and thus represent part of the degenerate ground states
of the Hamiltonian. In this way the MPS plays the role of a
generating state for a set of degenerate ground states of the
Hamiltonian, none of which has a MPS representation. The degree of
degeneracy of the ground states increases with system size, and each
state $|\psi_n\ra$ has a complicated structure, and can not be
represented as a matrix product, and thus the calculation of any of
its correlation functions, or even its normalization, is quite
difficult. However from the fact that the generating state
$|\psi(g)\ra$ is a matrix product state, we can determine such
correlations in closed form.

The structure of this paper is as follows: in section (\ref{MPS}) we
briefly review the matrix product formalism, with emphasis on the
symmetry properties of the state and the parent Hamiltonian, in
section (\ref{MODEL}) we consider three dimensional auxiliary
matrices and classify them according to symmetries of the states
which are constructed from them, namely symmetry with respect to
rotation around the z-axis and the discrete parity and spin-flip
symmetries. In this way we arrive at two specific forms of auxiliary
matrices and consequently two specific models. Sections
(\ref{modelI}) and (\ref{modelII}) are devoted to the detailed study
of the above two models. The paper ends with a conclusion and an
appendix.

\section{Symmetries of matrix product state and its parent
Hamiltonian}\label{MPS} From (\ref{psigeneral}) we see that the
collections of matrices $\{A_i\}$ and $\{\lambda UA_i U^{-1}\}$,
where $\lambda$ is a scalar, both define the same matrix product
state. This freedom allows us to study the symmetries of MPS. A MPS
will be symmetric under parity provided that we can find a matrix
$\Pi$ such that
$$\Pi A_m \Pi^{-1}\propto A_m^T$$ and invariant under spin flip
transformation, if we can find a matrix $\Omega$ such that $$\Omega
A_m \Omega^{-1} \propto A_{\overline{m}},$$ where here and
hereafter, $A_{\overline{m}}$ stands for $A_{-m}$. As for continuous
symmetries, consider a local symmetry operator $R$ acting on a site
as $R|i\ra=R_{ji}|j\ra$ where summation convention is being used.
$R$ is a $d$ dimensional unitary representation of the symmetry. A
global symmetry operator ${\cal R}:=R^{\otimes N}$ will then change
this state to another matrix product state
\begin{equation}\label{mpsPrime}
    \Psi_{i_1i_2\cdots i_N}\lo \Psi':=tr(A'_{i_1}A'_{i_2}\cdots
    A'_{i_N}),
\end{equation}
where
\begin{equation}\label{A'}
    A'_i:=R_{ij}A_j.
\end{equation}
A sufficient but not necessary condition for the state $|\Psi\ra$ to
be invariant under this symmetry is that there exist an operator
$U(R)$ such that
\begin{equation}\label{symm}
    R_{ij}A_j=U(R)A_iU^{-1}(R).
\end{equation}
Thus $R$ and $U(R)$ are two unitary representations of the symmetry,
respectively of dimensions $d$ and $D$. In case that $R$ is a
continuous symmetry operator with generators $T_a$, equation
(\ref{symm}), leads to
\begin{equation}\label{symmalg}
    (T_a)_{ij} A_j=[{\cal T}_a,A_i],
\end{equation}
where $T_a$ and ${\cal T}_a$ are the $d-$ and $D-$dimensional
representations of the Lie algebra of the symmetry. \\

A symmetric MPS need not be the ground state of a symmetric family
of Hamiltonians. To find the symmetric family of Hamiltonians we
should construct the Hamiltonian in a specific way. Let us first
review how the Hamiltonian is constructed in general.  From a matrix
product state, the reduced density matrix of $k$ consecutive sites
is given by
\begin{equation}\label{rhok}
    \rho_{i_1\cdots i_k, j_1\cdots j_k}=\frac{tr((A_{i_1}^*\cdots A_{i_k}^*\otimes A_{j_1}\cdots A_{j_k})E^{N-k})}{tr(E^N)}.
\end{equation}
The null space of this reduced density matrix includes the solutions
of the following system of equations
\begin{equation}\label{cc}
    \sum_{j_1,\cdots, j_k=1}^{d}c_{j_1\cdots
    j_k}A_{j_1}\cdots A_{j_k}=0.
\end{equation}
Given that the matrices $A_i$ are of size $D\times D$, there are
$D^2$ equations with $d^k$ unknowns. Since there can be at most
$D^2$ independent equations, there are at least $d^k-D^2$ solutions
for this system of equations. Thus for the density matrix of $k$
sites to have a null space it is sufficient that the following
inequality holds
\begin{equation}\label{dD}
    d^k\ >\ D^2.
\end{equation}
Let the null space of the reduced density matrix of $k$ consecutive
sites, denoted by ${\cal V}_k$,  be spanned by the orthogonal
vectors $|e_{\a}\ra, \ \ \ (\a=1, \cdots , s\geq d^k-D^2)$. Then we
can construct the local Hamiltonian acting on $k$ consecutive sites
as
\begin{equation}\label{h}
    h:=\sum_{\a=1}^s J_{\a} |e_{\a}\ra\la e_{\a}|,
\end{equation}
where $J_{\a}$'s are positive constants. These parameters together
with the parameters of the vectors $|e_\a\ra $ inherited from those
of the original matrices $A_i$, determine the total number of
coupling constants of the Hamiltonian.  If we call the embedding of
this local Hamiltonian into the sites $l$ to $l+k$ by $h_{l,l+k}$
then the full Hamiltonian on the chain is written as
\begin{equation}\label{H}
    H=\sum_{l=1}^N h_{l,l+k}.
\end{equation}
The state $|\Psi\ra$ is then a ground state of this Hamiltonian with
vanishing energy. The reason is as follows:
\begin{equation}\label{Hrho}
\la \Psi|H|\Psi\ra=tr(H|\Psi\ra\la\Psi|)=\sum_{l=1}^N
tr(h_{l,l+k}\rho_{l,l+k})=0,
\end{equation}
where $\rho_{l,k+l}$ is the reduced density matrix of sites $l$ to
$l+k$ and in the last equality we have used the fact that $h$ is
constructed from the null eigenvectors of $\rho$ for $k$ consecutive
sites. Given that $H$ is a positive operator, this proves the
assertion. For the Hamiltonian to have the symmetries of the ground
state, the basis vectors of the null space should be chosen so that
they transform into each other under the action of symmetries and
the couplings $J_{\a}$ should be chosen appropriately, see \cite{KM}
for a more detailed discussion of this point.
\section{Three dimensional auxiliary matrices}\label{MODEL} As is clear from (\ref{dD}) a sufficient
condition for the existence of a null space ${\cal V}_2$ for a
spin-one system is that the dimension of the matrices satisfy
$D^2<9$ which restricts $D$ to $1$ and $2$. The case of $D=1$ has
already been considered in \cite{kurman} outside the framework of
MPS formalism, and the case of $D=2$ has been worked out in
\cite{aklt} and \cite{zit1} as mentioned in the introduction.
However we should emphasize that this is a sufficient and not a
necessary condition and indeed we can take $D\geq3$ and still find a
non-empty null space ${\cal V}_2$, since the system of equations may
not all be independent of each other. \\

In this article we want to study in detail the case $D=3$ and
consider all the possible models which allow certain plausible
symmetries, i.e. rotational symmetry around the $z$ axis in spin
space, and symmetry under parity and spin flip operations, these are
the symmetries which have been taken into account in building
optimal ground states for various models \cite{fcs2, zit1, zit2,
aks2, aks3, zitladder, zit2dsquare, zit2dtriangular, akm1}.

So let us take 3-dimensional matrices $A_1, A_0$ and $A_{\1b}$ and
demand rotational symmetry around the $z$ axis in spin space.
According to (\ref{symmalg}), this is equivalent to the following
equations
\begin{equation}\label{}
    [S_z,A_1]=A_1, \h [S_z,A_0]=0, \h
    [S_z,A_{\overline{1}}]=-A_{\overline{1}},
\end{equation}
where $S_z = diagonal (1,0,-1)$. The immediate solution of these
equations is
\begin{equation}
A_1=\left(\begin{array}{c c c}
0&a&0\\
0&0&b\\
0&0&0\end{array}\right)\hskip 1cm A_0=\left(\begin{array}{c c c}
g&0&0\\
0&h&0\\
0&0&i\end{array}\right)\hskip 1cm
A_{\overline{1}}=\left(\begin{array}{c c c}
0&0&0\\
c&0&0\\
0&d&0\end{array}\right),
\end{equation}
where $a, b, c, d, g, h$ and $i$ are real parameters. By a
transformation $A_i \lo S A_i S^{-1}$ where $S = diagonal (1, a,
ab)$ we can set the parameters of $A_1$ equal to 1.  Symmetry under
parity now requires that there is a matrix $\Pi$ such that
\begin{equation}\label{}
\Pi A_m \Pi^{-1}= A^T_{m}.
\end{equation}
A straightforward calculation gives
\begin{equation}\label{}
    \Pi=\left(\begin{array}{ccc} 0 & 0 & 1 \\ 0 & 1 & 0 \\ 1 & 0 & 0 \end{array}\right)
\end{equation}
and
\begin{equation}\label{}
 A_1=\left(\begin{array}{ccc}0&1&0\\ 0 & 0 & 1 \\ 0 & 0 & 0\end{array}\right),\ \ \
    A_0=\left(\begin{array}{ccc}g&0&0\\ 0 & h & 0 \\ 0 & 0 & g\end{array}\right),\ \ \
    A_{\1b}=\left(\begin{array}{ccc}0&0&0\\ c & 0 & 0 \\ 0 & c &
     0\end{array}\right).
\end{equation}

Finally we come to the symmetry under spin flip $|m\ra \lo
|\overline{m}\ra$. It is readily seen that with these matrices the
spin-flip symmetry is automatic, namely we have
\begin{equation}\label{}
    \Omega A_m\Omega^{-1}=A_{\overline{m}},
\end{equation}
in which
\begin{equation}\label{}
    \Omega=\left(\begin{array}{ccc} 0 & 0 & 1 \\ 0 & {c} & 0 \\ {c^2} & 0 & 0
    \end{array}\right).
\end{equation}

In order to guarantee that the matrix product state constructed in
this way is the ground state of a Hamiltonian with nearest neighbor
interaction, we consider the equation

\begin{equation}\label{}
    \sum_{i,j=1,0,\overline{1}}c_{ij}A_iA_j=0,
\end{equation}
which can be re-written as a matrix equation for the coefficients
$c_{ij}$ in the form

\begin{equation}\label{}
    \sum_{i,j}M_{kl,ij}c_{ij}=0.
\end{equation}
To have a solution we set
\begin{equation}\label{}
    det(M)=0.
\end{equation}
The determinant of $M$ is readily calculated from its explicit form
and is given by
\begin{equation}\label{}
    det(M)=(g^2-h^2)^2(2g^2-h^2)c^4.
\end{equation}
The vanishing of the determinant puts constrains on the parameters,
namely we should have either $c=0$, or $h=\pm g$ or $h=\pm
\sqrt{2}g$, each choice leading to a different exactly solvable
model. We omit the case $c=0$ since it leads to the condition
$A_{\overline{1}}=0$ and hence reduces the model to an effectively
two-state model, moreover in this case, spin-flip symmetry is lost
due to the non-existence of an invertible matrix $\Omega$. Also it
turns out that the other models with minus signs are equivalent to
models with plus signs, see appendix {\bf A} for a demonstration of
this fact. So we are left with two different models which we label
accordingly as model I (when $h=\sqrt{2}g$) and model II
(when $h=g$) and study them separately in subsequent sections. \\

Before proceeding to the models, we need to clarify a point about
the number of parameters. Throughout our analysis we take $N$, the
size of the lattice to be an even number. It appears that we have
two continuous parameters in the matrix product states, namely $g$
and $c$. None of these parameters can be gauged away by similarity
transformations or scaling of the auxiliary matrices. However the
MPS depends on only one parameter. To see this, let us expand the
MPS in terms of the states $|\psi_n\ra$ which are defined to be
linear superposition of all states which have $n$ $0$'s. Note that
for an even $N$, $n$ will also have to be even. Since in the space
of one site, the operators $A_1$ and $A_{\1b}$ act as raising and
lowering operators, the trace of any string of operators $A_0, A_1$
and $A_{\1b}$ is non-vanishing only if this string contains an equal
number of $A_1$ and $A_{\1b}$. Thus any state $|\psi_n\ra$ comes
with a coefficient $g^nc^{\frac{N-n}{2}}$. Consequently for the
un-normalized MPS we have
\begin{equation}\label{psigggg}
    |\psi(g,c)\ra=\sum_{n=0}^N g^nc^{\frac{N-n}{2}}|\psi_n\ra \equiv
    c^\frac{N}{2}\sum_{n=0}^N (\frac{g^2}{c})^{\frac{n}{2}}|\psi_n\ra.
\end{equation}
Thus the normalized state and all the correlation functions will
depend on only one single parameter, namely $\frac{g^2}{c}$. For
this reason we can put $c=1$ and so the MPS will depend on only one
single parameter $g$.

\section{Model I}\label{modelI}
In this section we study in detail model I. The auxiliary matrices
are
\begin{equation}\label{matricesI}
 A_1=\left(\begin{array}{ccc}0&1&0\\ 0 & 0 & 1 \\ 0 & 0 & 0\end{array}\right),\ \ \
    A_0=g\left(\begin{array}{ccc}1&0&0\\ 0 & \sqrt{2} & 0 \\ 0 & 0 & 1\end{array}\right),\ \ \
    A_{\1b}=\left(\begin{array}{ccc}0&0&0\\ 1 & 0 & 0 \\ 0 & 1 &
     0\end{array}\right).
\end{equation}
First we derive the one-parameter family of parent Hamiltonians and
then calculate the one and the two-point functions.
\subsection{The Hamiltonian}Here we have $h=\sqrt{2}g$ and the null space ${\cal V}_2$
is spanned by one single vector \be\label{eI}
|e\ra=|00\ra-g^2|1,\overline{1}\ra-g^2|\overline{1},1\ra. \ee We
take the local Hamiltonian as $h_I=\frac{1}{(1-g^2)^2}|e\ra\la e|$
from which the full Hamiltonian turns out to be
\begin{equation}\label{HI}
    H_{I}(u)=(1-u^2)N+\sum_i S_{z,i}^2 S_{z,i+1}^2 - 2(1+2u)S_{z,i}^2+ u^2
    ({\bf S}_i\cdot {\bf S}_{i+1})^2 + u\{ {\bf S}_i\cdot {\bf
    S}_{i+1},S_{z,i}S_{z,i+1}\},
\end{equation}
where $u:=\frac{g^2}{1-g^2}$. When $g=u=0$, the Hamiltonian turns
into
\begin{equation}\label{Hu0}
H_1=\sum_{i}(S^2_{z,i}-1)(S^2_{z,i+1}-1).
\end{equation}
In this limit, since $A_0\rightarrow 0$, the MPS becomes an
expansion of states consisting only of $1$ and $\overline{1}$. Such
a state is clearly the ground state of the Hamiltonian (\ref{Hu0}),
however, the Hamiltonian (\ref{Hu0}) has a highly degenerate ground
state, which is not captured by the MPS in this limit. In fact, any
basis state in which no two $0$'s are adjacent is a ground state of
this Hamiltonian, with energy $E_0=0$. The number of ground states
of $H_1$ is equal to $tr(A^N)$, where $N$ is the system size and $A$
is the adjacency matrix in which allowed adjacent configurations are
designated by $1$ and disallowed configurations by $0$. In $H_1$ the
only configuration which lifts the local energy from $0$ to $1$ is
that of two adjacent zeros, so in the basis which we have chosen,
$A=\left(\begin{array}{ccc} 1 & 1 & 1 \\ 1 & 0 & 1
\\ 1 & 1 & 1 \end{array}\right)$. For large $N$, we will have $tr(A)^N\approx
(1+\sqrt{3})^N$, where $1+\sqrt{3}$ denote the largest eigenvalue of
$A$.\\

When $g\rightarrow 1$ (or $u\rightarrow \infty$), $H_I(u)$ turns,
modulo a multiplicative coefficient, into the following Hamiltonian,
\begin{equation}\label{Huinfinity}
H_2=\sum_{i}[({\bf S}_i\cdot{\bf S}_{i+1})^2-1].
\end{equation}
This is a simple and interesting Hamiltonian and thus our result
implies that its ground state is of the form of an MPS, with
matrices given by (\ref{matricesI}), for $g=1$.

Note that in the limit $g=1$, the null eigenvector (\ref{eI})
becomes a singlet, the spin-0 representation of angular momentum,
implying that the Hamiltonian should be a scalar which conforms with
the form of the Hamiltonian (\ref{Huinfinity}). However the matrices
(\ref{matricesI}) (for $g=1$) do not transform as a spherical tensor
operator under angular momentum, that is the following relations are
not satisfied as required from (\ref{symmalg}):
\begin{eqnarray}
  [S_z,A_m] &=& mA_m \cr
  [S_+,A_m] &=& \sqrt{2-m(m+1)}A_{m+1} \cr
  [S_-,A_m] &=& \sqrt{2-m(m-1)}A_{m-1},
  \end{eqnarray}
where $S_z$, and $S_{\pm}$ are the three dimensional representation
of $su(2)$. This means that the Hamiltonian is symmetric under the
full rotation group, but the ground state, breaks this symmetry.
Therefore other degenerate ground states can be constructed by
applying rotation operators to this state. However the actual
degeneracy is much larger and it grows exponentially with system
size as $(\frac{3+\sqrt{5}}{2})^N$ \cite{klumper93}.\\

We should stress that the Hamiltonian $H_2$ corresponds to a
particular point in the class of bilinear-biquadratic spin-1 chains
with the Hamiltonian \be H(\theta):= \sum_{i=1}^N \left[\cos\theta \
{\bf S}_i\cdot {\bf S}_{i+1} +\sin\theta\ ({\bf S}_i\cdot {\bf
S}_{i+1})^2\right]\ee which has extensively been studied by various
methods \cite{bb1,bb2,bb3,bb4,bb5,bb6}.

\subsection{Correlation functions} The correlation functions of this model are determined
after lengthy but straightforward calculations starting from
(\ref{corrgeneral}) and (\ref{Eo}). In the thermodynamic limit, the
results are as follows:
\begin{equation}\label{correlationsI}
    \la S_{z,i}\ra=\la S_{x,i}\ra=0.
\end{equation}
Thus in the ground state, there is no magnetization. Note that due
to rotational invariance in the $x-y$ plane of spin space, in all
the correlation functions below, we can change $x$ to $y$ or any
other direction in the $x-y$ plane.   To describe the other
correlation functions, let us introduce the parameter
$\gamma:=\sqrt{g^4+8}$. Then we have
\begin{equation}\label{correlationsI}
    \la S^2_{z,i}\ra=\frac{8}{\gamma(3g^2+\gamma)},\h \la S^2_{x,i}\ra=
    \frac{g^2(3\gamma+g^2)+4}{\gamma(3g^2+\gamma)}.
\end{equation}
\begin{figure}[t]
 \centering
   \includegraphics[width=8cm,height=6cm,angle=0]{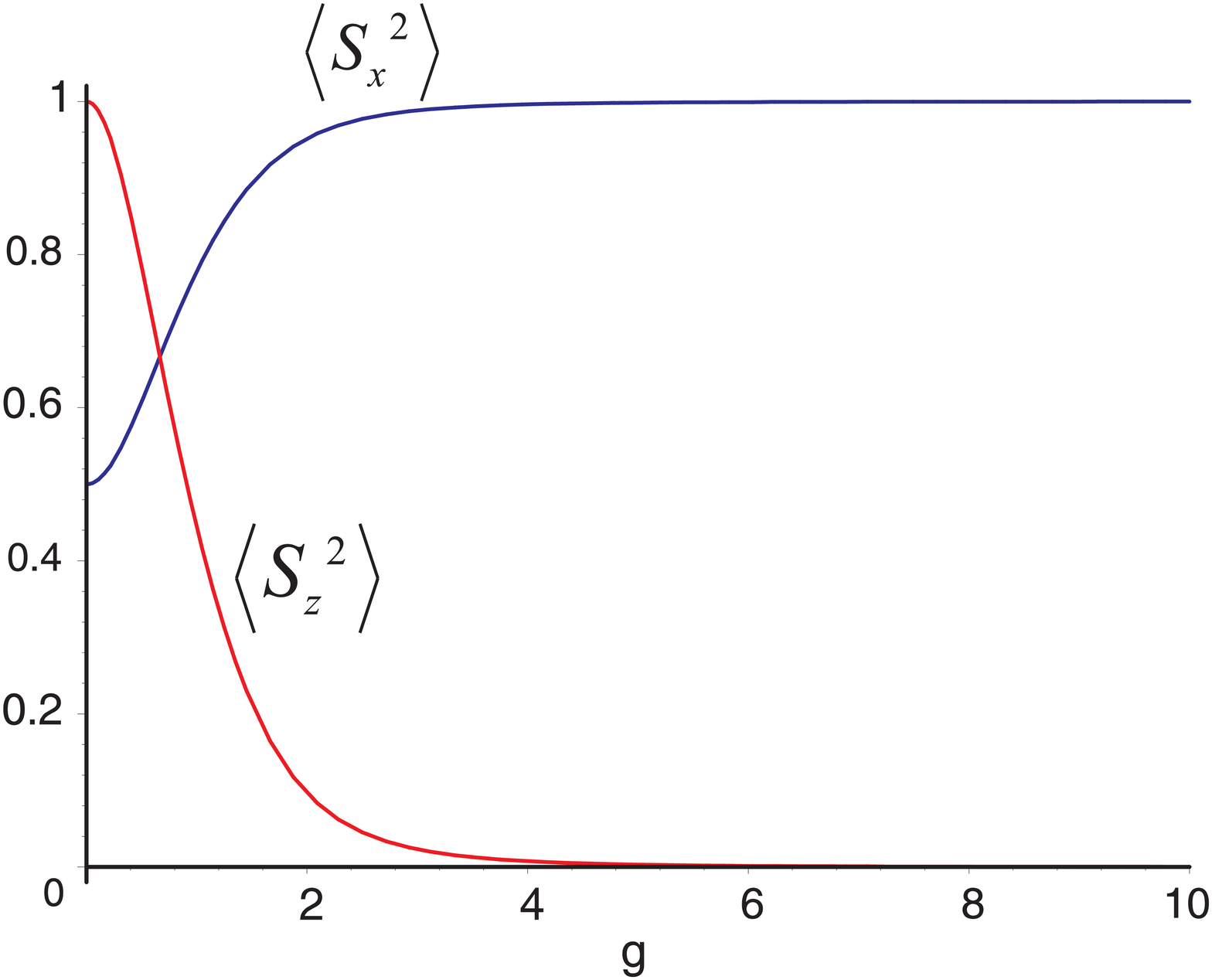}
   \caption{The average values $\la S_z^2\ra$ and $\la S_x^2\ra$ for model I.
    In the limit $g\rightarrow \infty (u\rightarrow -1)$}, the spins
    lie in the $x-y$ plane.
    \label{Squarecorrelations}
\end{figure}

\begin{figure}[t]
 \centering
   \includegraphics[width=12cm,height=6cm,angle=0]{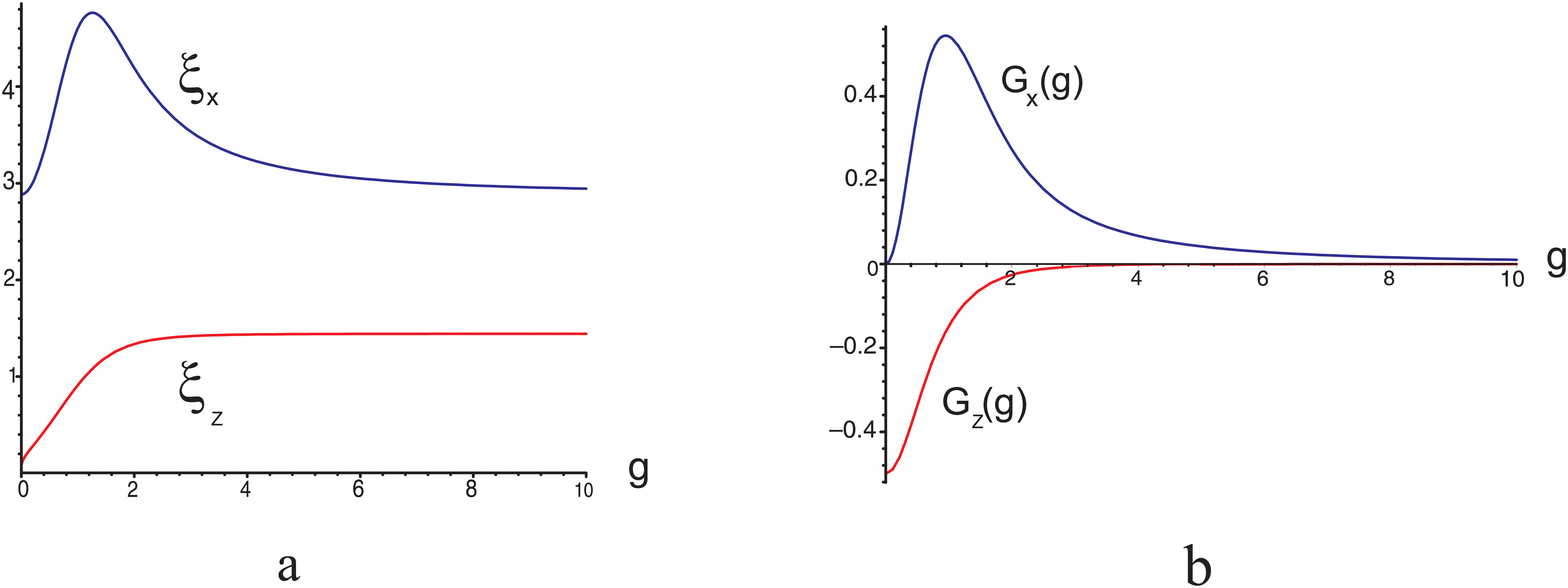}
   \caption{The correlation functions for longitudinal and transverse components of spins,
   in model I; a) Correlation lengths, b)
    the magnitudes of correlations. }
   \label{correlations}
\end{figure}
Figure (\ref{Squarecorrelations}) shows the plot of $\la S^2_{z,i}\ra$  and $\la
S^2_{x,i}\ra$ as a function of the parameter $g$.  In the limit
$g\rightarrow 0$, we have $\la  S^2_{z,i}\ra\rightarrow 1$, implying
that there is no $0$ in the expansion of the state. This is in
accord with our picture of the MPS, since in this limit
$A_0\rightarrow 0$.
 Finally
for the two point correlations of longitudinal and transverse
components of spins we find
\begin{eqnarray}\label{correlationsI}
    \la S_{z,1}S_{z,r}\ra&=&G_{||}(g)\ e^{-(r-2)/\xi_{||}},\cr
\la S_{x,1}S_{x,r}\ra&=& G_{\perp}(g)\ e^{-(r-2)/\xi_{\perp}},
\end{eqnarray}
where the magnitude of correlations are given by
\begin{equation}
    G_{||}(g)= -\frac{4(g^2+\gamma)}{\gamma(3g^2+\gamma)^2},\h
    G_{\perp}(g) =
    \frac{8g^2(g^2+\sqrt{2}+\gamma)^2}{\gamma(3g^2+\gamma)^2(g^2+\gamma)},
\end{equation}
and the longitudinal and transverse correlations are given by
\begin{equation}
    \xi_{||}= \frac{1}{\ln (\frac{3g^2+\gamma}{2g^2})},\h
    \xi_{\perp} = \frac{1}{\ln(\frac{3g^2+\gamma}{2+2\sqrt{2}g^2})}.
\end{equation}
These are plotted in figure (\ref{correlations}). Note that
$G_{||}(g)=\la S_{z,1}S_{z,2}\ra$ and $G_{\perp}(g)=\la
S_{x,1}S_{x,2}\ra$.  In the limit $g\lo 0 $, where $H_I$ becomes an
Ising-like Hamiltonian, the above equations show that transverse
correlations vanish, and longitudinal correlations approach the
value $-1/2$. However we should note that in this limit, the ground
state is highly degenerate. In fact as stated above, any basis state
in which there are no two adjacent $0$'s is a ground state of $H_1$.
However the MPS does not capture this degeneracy, but is only one of
the many ground states. In the limit $g\rightarrow 1$, where the
Hamiltonian turns into (\ref{Huinfinity}), we have $\la
S^2_{z,i}\ra=\frac{4}{9}$ and $\la S^2_{x,i}\ra=\frac{7}{9}$, and
the correlation lengths tend to $\xi_{||}=\frac{1}{\ln (3)}=0.910$
and $\xi_{\perp}=\frac{1}{\ln (3/(1+\sqrt{2}))}=4.603$.

\section{Model II}\label{modelII}
For this model, the auxiliary matrices are
\begin{equation}\label{matriceII}
 A_1=\left(\begin{array}{ccc}0&1&0\\ 0 & 0 & 1 \\ 0 & 0 & 0\end{array}\right),\ \ \
    A_0=g\left(\begin{array}{ccc}1&0&0\\ 0 & 1 & 0 \\ 0 & 0 & 1\end{array}\right),\ \ \
    A_{\1b}=\left(\begin{array}{ccc}0&0&0\\ 1 & 0 & 0 \\ 0 & 1 &
     0\end{array}\right).
\end{equation}

\subsection{The Hamiltonian}
\begin{figure}[t]
 \centering
   \includegraphics[width=9.5cm,height=8cm]{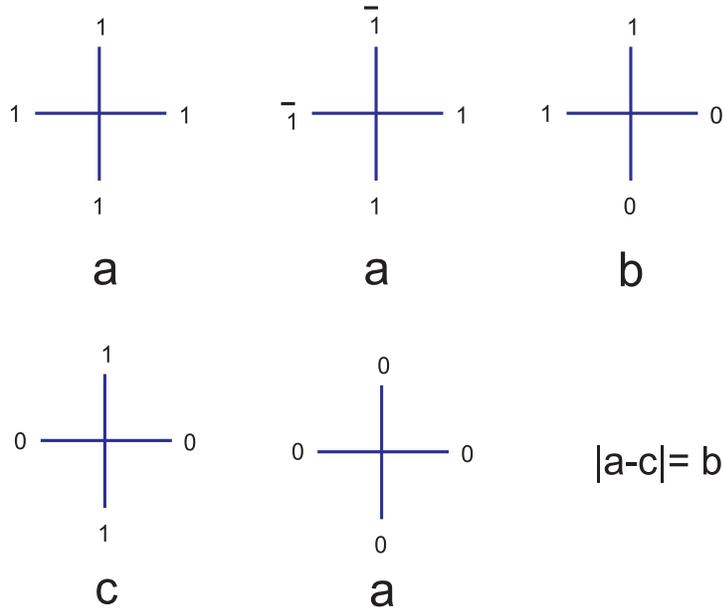}
   \caption{The independent Boltzmann weights of the two
dimensinal vertex model corresponding to the quantum spin chain
(\ref{oldH}).}
   \label{Boltzmann}
\end{figure}

The null space ${\cal V}_2$ is spanned by the following two vectors:
\begin{eqnarray}\label{e}
|e_1\ra=\frac{1}{\sqrt{2}}(|01\ra-|10\ra),\cr
|e_2\ra=\frac{1}{\sqrt{2}}(|0\overline{1}\ra-|\overline{1}0\ra).
\end{eqnarray}
These vectors are eigenvectors of the local two-site $S_z$ operator,
are invariant under parity and transform into each other under spin
flip transformation. Therefore if we take the local symmetric
Hamiltonian as
\begin{equation}\label{h}
h_{II}=|e_1\ra\la e_1|+|e_2\ra\la e_2|,
\end{equation}
where we have set a total multiplicative constant equal to unity,
the final total Hamiltonian is spin-flip and parity invariant and
moreover commutes with the third component of spin, i.e.
$[h_{II},S_z]=0$. Its explicit form in terms of local spin operators
can be determined after some algebra:
\begin{equation}\label{oldH}
H_{II}=\sum_{i=1}^{N}2S^2_{z,i}- \{ {\bf S}_i.{\bf
S}_{i+1},S_{z,i}S_{z,i+1}\} - {\bf S}_i.{\bf
S}_{i+1}+S_{z,i}S_{z,i+1}.
\end{equation}
This Hamiltonian was first found in \cite{KlumperRep1}. The history
is the following: The exhaustive solutions of the Yang-Baxter
equation, corresponding to a three-state 19-vertex model on a square
lattice were first found in \cite{IdzumiRep1}. These solutions
reproduced many of the already known exactly solvable vertex models
in addition to four new models. These models, labeled I, II, III and
IV in \cite{KlumperRep1}, were then studied in detail in
\cite{KlumperRep1}, where the thermodynamic properties of these new
models, including the partition function and correlation lengths
were derived. Two of these models, namely models I and II, however
allowed a more complete solution (due to the so called crossing
symmetry of the $R$ matrix, the solution of the Yang-Baxter
equation) which allowed the exact determination of the ground state
energy per site. However the other two models, models III and IV,
lacked this symmetry, and no exact solution for the ground state
energy was given. It could however be established that such models
can be mapped to 6-vertex models, i.e. two state models with 6
allowed configurations. The Hamiltonian (\ref{oldH}) corresponds in
fact to the Hamiltonian of model III in \cite{KlumperRep1}, for
$\Delta=1$, where $\Delta$ is a particular combination of Boltzmann
weights.  For its definition and the Boltzmann weights see
\cite{KlumperRep1}. According to \cite{KlumperRep1}, the three-state
19-vertex model corresponding to this spin chain is defined by the
Boltzmann weights shown in figure (\ref{Boltzmann}).

Note that each Boltzmann weight $R^{\mu,\a}_{\nu,\beta}$ corresponds
to the spin labels, $\mu,\nu,\a,\beta$ on left, right, bottom and
top links of a vertex and the $R$ matrix and correspondingly the
Boltzmann weights have the following symmetries (with
$\overline{\a}:=-\a, etc):$

\begin{equation}\label{BoltzmannSymmetries}
    R^{\mu,\a}_{\nu,\beta}=R^{\a,\mu}_{\beta,\nu}=
    R^{\nu,\beta}_{\mu,\a}=R^{\overline{\mu},\overline{\a}}_{\overline{\nu},\overline{\beta}}.
\end{equation}

The exact correspondence between the Hamiltonian (\ref{oldH}) and
that of model III of \cite{KlumperRep1} is as follows:
\begin{equation}\label{corr}
    H_{II}=H_{6-{\rm vertex}} + N,
\end{equation}
where $N$ is the number of sites of the chain. This correspondence
allows to determine exactly the ground state energy of the
corresponding 6-vertex model, for the range of values of Boltzmann
weights of the 6-vertex model shown in figure (\ref{Boltzmann}).
Since the ground state energy of $H$ is zero, we find the ground
state energy of $H_{6-{\rm vertex}}$ to be $E_0=-N$, giving an
energy per site equal to $e_0:=\frac{E_0}{N}=-1$.\\
\subsection{The explicit form of the ground states}
From the form of the local Hamiltonian one finds that any state
which comprises of only $1$'s and $\overline{1}$'s, with no $0$'s,
is a ground state of this Hamiltonian. The number of such states is
$2^{N}$, where $N$ is the size of the system and their explicit form
is $|s_1,s_2,\cdots s_{N}\ra$, where $s_i\in \{1,\overline{1}\}$.
However there are other non-trivial ground states, those which
contain also a number of $0$'s. The interesting point is that these
kinds of ground states, are nicely captured by the MPS
(\ref{matriceII}). We can find $N$ such degenerate states. To see
this we note that MPS (\ref{matriceII}) depends on a continuous
parameter $g$ while the Hamiltonian does not. This dependence is
genuine and can not be gauged away by similarity transformation
$A_m\lo UA_mU^{-1}$. Let us expand the MPS in terms of powers of $g$
in the form
\begin{equation}\label{}
    |\psi(g)\ra = \sum_{n=0}^{N} g^n|\psi_{n}\ra,
\end{equation}
where
\begin{equation}\label{phin}
    |\psi_n\ra = \sum'_{i_1,i_2,\cdots i_N} tr(A_{i_1}A_{i_2}\cdots
    A_{i_N})|i_1,i_2,\cdots i_N\ra,
\end{equation}
and $\sum'$ implies that in each term only $n$ $A_0$'s exist.\\

In view of (\ref{matriceII}) and the definition of MPS
(\ref{psimps}) the powers of $g$ enumerate the number of $0$'s in
each state and so each $|\psi_n\ra$ is a superposition of states
each of which has exactly $n$ local $0$'s and an equal number of
$1$'s and $\overline{1}$'s.  Note that since we have taken $N$ to be
even, this
implies that $n$ is also an even number.\\

To determine the explicit form of each state $|\psi_n\ra$, consider
the form of the matrices $A_0, A_{1}$ and $A_{\overline{1}}$ in
(\ref{matriceII}). With
$$|1\ra:=\left(\begin{array}{c} 1 \\ 0 \\ 0 \end{array}\right),
\h |0\ra:=\left(\begin{array}{c} 0 \\ 1 \\ 0
\end{array}\right) \h |\overline{1}\ra:=\left(\begin{array}{c} 0 \\ 0 \\ 1
\end{array}\right),$$
we have
\begin{equation}\label{MGmatrices}
A_1=|1\ra\la 0 |+|0\ra\la \overline{1}|, \h A_0 = gI, \h
A_{\overline{1}}=|\overline{1}\ra\la 0 |+|0\ra\la 1|,
\end{equation}
or more compactly
\begin{equation}\label{CompactMGmatrices}
A_m=|m\ra\la 0 |+|0\ra\la \overline{m}|, \h A_0 = gI,
\end{equation}
where $m=1,\overline{1}$ and $I$ is the identity matrix. The product
of any string of matrices $A_m$, $m=1,\overline{1}$ has a simple
structure. One can easily show that
\begin{eqnarray}\label{Am}
    A_{m_1}A_{m_2}\cdots
    A_{m_{2K}}&=&\delta_{m_1,\overline{m_2}}\delta_{m_3,\overline{m_4}}
    \cdots \delta_{m_{2K-1},\overline{m_{2K}}}|0\ra\la 0| \cr &+& \delta_{\overline{m_2},m_3}
    \delta_{\overline{m_4},m_5}\cdots \delta_{\overline{m_{2K-2}},m_{2K-1}}|m_{1}\ra\la
    \overline{m_{2k}}|.
\end{eqnarray}

\begin{figure}[t]
 \centering
   \includegraphics[width=10cm,height=2.5cm]{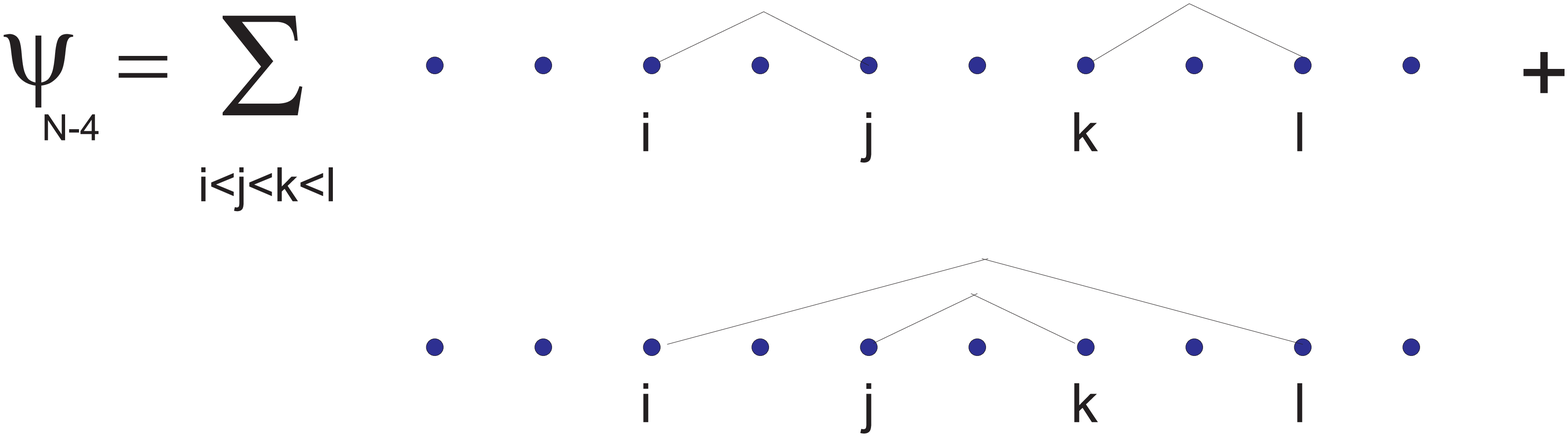}
   \caption{The state $|\psi_{N-4}\ra$. All the dots are at the state $|0\ra$, while those which are connected
   are at the state $|1,\overline{1}\ra + |\overline{1},1\ra$.
}
   \label{StatesN-4}
\end{figure}

\begin{figure}[t]
 \centering
   \includegraphics[width=10cm,height=1.5cm]{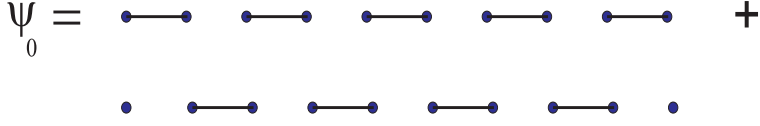}\vspace{0.5cm}
   \caption{The state $|\psi_0\ra$ has a Majumdar-Ghosh like
   structure. Note that a line here means only an entangled state
   $|1,\overline{1}\ra+|\overline{1},1\ra$ and not a spin singlet.
   }
   \label{States0}
\end{figure}

This allows us to determine the explicit form of any of the states
$|\psi_n\ra$. Consider for example $|\psi_{N}\ra$, where there is no
$1$ or $\overline{1}$. This is a simple state
$|\psi_{N}\ra=3|0,0,\cdots 0\ra$, where the factor $3$ comes from
taking the trace of the identity matrix, related to the product of
all the $A_0$ matrices. The next state is $|\psi_{N-2}\ra$ in which
two of the zeros have been replaces by $1$ and $\overline{1}$. Using
(\ref{Am}) we find that this is a state of the form

\begin{equation}\label{psi2}
    |\psi_{N-2}\ra=2\sum_{m<n}(|0,\cdots 0, 1_{m},0,\cdots, 0 , \overline{1}_n,0,\cdots 0\ra+
    |0,\cdots 0, \overline{1}_m,0,\cdots, 0 , 1_n,0,\cdots
    0\ra),
\end{equation}
where the two non-zero spins occur at sites $m$ and $n$
respectively. Let us define
$|\a_{ij}\ra:=|1_i,\overline{1}_j\ra+|\overline{1}_i,1_j\ra$ where
the indices denote the sites of the lattice, and all other sites are
occupied by $0$. Then $|\psi_{N-2}\ra=2\sum_{i<j}|\a_{ij}\ra$. In
view of (\ref{Am}), we find the structure of all other states
$|\psi_{n}\ra$. For example we have
\begin{equation}\label{psi2}
    |\psi_{N-4}\ra=\sum_{i<j< k<l}(|\a_{ij}\ra|\a_{kl}\ra+|\a_{jk}\a_{li}\ra,
\end{equation}
which is pictorially depicted in figure (\ref{StatesN-4}) and
\begin{equation}
  |\psi_{N-6}\ra = \sum_{i<j<k<l<m<n} \left(|\a\ra_{ij}|\a\ra_{kl}|\a\ra_{mn} +
  |\a\ra_{ni}|\a\ra_{jk}|\a\ra_{lm}\right).
\end{equation}
Finally we find $\psi_{0}$, in which there are no zeros, and has a
dimmerized or Majumdar-Ghosh like structure, namely
\begin{equation}
  |\psi_{0}\ra =
  |\a_{12}\ra|\a_{34}\ra\cdots|\a_{N-1,N}\ra+|\a_{23}\ra|\a_{45}\ra\cdots|\a_{N,1}\ra,
\end{equation}
which is depicted in figure (\ref{States0}).

\vspace{0.5cm}

\subsection{Correlation functions}
We have found that the number of degenerate ground states of model
II, for a chain of size $N$, is at least $2^{N}+N$. Of these, the
$2^N$ states $|s_1,s_2,\cdots s_N\ra$, where $s_i\in
\{1,\overline{1}\}$ are uncorrelated, even for finite-size systems.
The other $N$ states are non-trivial and have MP representations.
Here we calculate the correlation functions for the other types of
states which are correlated. Were it not for the matrix product
formalism, such calculation could be very difficult. As a first step
let us determine the normalization of the states $|\psi_n\ra$, and
then proceed to the calculation of correlation functions of various
operators.

Let us first fix our notation for matrix product operators, using
their general definition (\ref{Eo}). We have
\begin{eqnarray}\label{mpsop}
    E_{S_z}&=&A_1\otimes A_1-A_{\overline{1}}\otimes
    A_{\overline{1}}=:U,\cr
    E_{S_z^2}&=&A_1\otimes A_1+A_{\overline{1}}\otimes
    A_{\overline{1}}=V,\cr
    E\ &=&A_1\otimes A_1+A_{\overline{1}}\otimes
    A_{\overline{1}}+A_0\otimes A_0=V+g^2I.
    \end{eqnarray}
Let us start by finding the normalization of the states
$|\psi_n\ra$. It is obvious that for $m\ne n$, $\la
\psi_m|\psi_n\ra=0,$ since these two states, have different number
of zeros in their expansion. Therefore
\begin{equation}\label{psiext1}
    \la \psi(g)|\psi(g)\ra=\sum_{n=0}^{N}g^{2n}\la
    \psi_n|\psi_n\ra.
\end{equation}
The left hand side is found from (\ref{norm}) to be $$ \la
\psi(g)|\psi(g)\ra=tr(E^N)=tr(V+g^2I)^N =\sum_{n=0}^N g^{2n} \left(\begin{array}{c}N\\
n\end{array}\right) tr(V^{N-n}).$$ Comparing with the previous
formula, we find
\begin{equation}\label{psin}
    \la \psi_n|\psi_n\ra =\left(\begin{array}{c}N\\
n\end{array}\right)tr(V^{N-n}).
\end{equation}
\subsubsection{One-point functions}
In the thermodynamic limit $N\lo \infty$ (n=finite), the traces are
simplified by taking only the largest eigenvalues, but before going
to this limit, let us derive closed formulas for some of the
correlation functions. From the nature of the states $|\psi_n\ra,$
it is obvious that
\begin{equation}\label{}
    \la \psi_n|S_z|\psi_n\ra=\la \psi_n|S_a|\psi_n\ra=0,
\end{equation}
where $a$ is any direction perpendicular to the $z$ axis. To obtain
the average of $S_z^2$, we note again that for $m\ne n$, $\la
\psi_n|S_z^2|\psi_m\ra=0$ and hence
\begin{equation}\label{Z2n}
    \la \psi(g)|S_z^2|\psi(g)\ra=\sum_{n=0}^N g^{2n}\la
    \psi_n|S_z^2|\psi_n\ra.
\end{equation}
The left hand side is obtained from (\ref{corrgeneral}):
\begin{equation}\label{}
    \la \psi(g)|S_z^2|\psi(g)\ra = tr(E_{S^2_z}E^{N-1})=tr(V(V+g^2I)^{N-1})= \sum_{n=0}^{N-1} g^{2n}
    \left(\begin{array}{c}N-1\\
n\end{array}\right) tr(V^{N-n}),
\end{equation}
which in view of (\ref{Z2n}) gives
\begin{equation}\label{}
    \frac{\la \psi_n|S_z^2|\psi_n\ra}{\la\psi_n|\psi_n\ra} = \frac{\left(\begin{array}{c}N-1\\
n\end{array}\right)}{\left(\begin{array}{c}N\\
n\end{array}\right)}=\frac{N-n}{N},
\end{equation}
as expected, since the operator $S_{z}^2$ counts the average number
of $1$'s or $\overline{1}$'s in the state. From $\la
S_x^2+S_y^2+S_z^2\ra=2 $ and the rotational symmetry around the $z$
axis, we find for any unit vector $a$ in the $x-y$ plane
\begin{equation}\label{}
    \frac{\la
    \psi_n|S_a^2|\psi_n\ra}{\la\psi_n|\psi_n\ra}=\frac{N+n}{2N}.
\end{equation}
\subsubsection{Two-point functions}
Let us now derive the two point correlation functions,
$\la\psi_n|S_{z,1}S_{z,r}|\psi_n\ra$. We have
$$ \la\psi(g)|S_{z,1}S_{z,r}|\psi(g)\ra =
tr(E_{S_z}E^{r-2}E_{S_z}E^{N-r})=tr(U(V+g^2I)^{r-2}U(V+g^2I)^{N-r})
$$ Expanding the right hand side and using the diagonal
property of this two-point function, namely that for $m\ne n$, $\la
\psi_n|S_{z,1}S_{z,r}|\psi_m\ra=0,$ we find
\begin{equation}\label{z1zrn}
    \frac{\la \psi_n|S_{z,1}S_{z,r}|\psi_n\ra}{\la
    \psi_n|\psi_n\ra}=\frac{\sum_{k=0}^{r-2}\left(\begin{array}{c}r-2\\
k\end{array}\right)\left(\begin{array}{c}N-r\\
n-k\end{array}\right)tr(UV^{r-2-k}UV^{N-r-n+k})}{\left(\begin{array}{c}N\\
n\end{array}\right)tr(V^{N-n})}.
\end{equation}

The expansion of $\la\psi(g)|S^2_{z,1}S^2_{z,r}|\psi(g)\ra$ gives a
simple result, since in this case, the matrix product operator
$E_{S^2_z}=V$ commutes with $E$, that is

\begin{eqnarray}\label{}
    \la\psi(g)|S^2_{z,1}S^2_{z,r}|\psi(g)\ra&=&
    tr(V(V+g^2I)^{r-2}V(V+g^2I)^{N-r})=tr(V^2(V+g^2I)^{N-2})\cr
    &=& \sum_{n=0}^{N-2}\left(\begin{array}{c}N-2\\
n\end{array}\right) g^{2n}tr(V^{N-n}),
\end{eqnarray}
giving the final distance-independent result
\begin{equation}\label{}
    \frac{\la \psi_n|S^2_{z,1}S^2_{z,r}|\psi_n\ra}{\la
    \psi_n|\psi_n\ra}=\frac{\left(\begin{array}{c}N-2\\
n\end{array}\right)}{\left(\begin{array}{c}N\\
n\end{array}\right)}=\frac{(N-n)(N-n-1)}{N(N-1)}.
\end{equation}

Finally we come to the two-point function $\la S_{x,1}S_{x,r}\ra$
between transverse components of spins. Here we encounter an
essential difference with the previous cases, in that the operator
$S_{x}$ is not diagonal between the states $|\psi_n\ra$, i.e. $\la
\psi_n|S_{x,1}S_{x,r}|\psi_m\ra\ne 0$ for $m\ne 0$. To circumvent
this problem, we do the following expansion, for $g$ and $h$ two
arbitrary variables:
\begin{equation}\label{SxSx}
    \la \psi(g)|S_{x,1}S_{x,r}|\psi(h)\ra=\sum_{n,m=0}^N g^{n}h^m\la
    \psi_n|S_{x,1}S_{x,r}|\psi_m\ra.
\end{equation}
To calculate the left hand side within the matrix product formalism,
we should slightly adapt equations (\ref{norm}), (\ref{corrgeneral}) and (\ref{Eo}) to conform to the
present situation. Let $|\phi\ra$ and $|\psi\ra$ be two matrix
product states on the same periodic lattice of size $N$, defined by
matrices $\{A_i\}$ and $\{B_i\} $ respectively. Then by following
the steps which led to (\ref{norm}) and (\ref{corrgeneral}) we find
\begin{equation}\label{product}
    \la \phi|\psi\ra=tr(\tilde{E}^N),
\end{equation}
where $\tilde{E}=\sum_{i}A_i^*\otimes B_i$. The one point function
of an observable $O$ will be given by
\begin{equation}\label{product}
    \la \phi|O|\psi\ra=tr(\tilde{E}_O\tilde{E}^{N-1}),
\end{equation}
where
\begin{equation}\label{}
    \tilde{E}_O=\sum_{i,j} A_i^* \otimes B_j \la i|O|j\ra.
\end{equation}

We now use the above formulas to calculate the left hand side of (\ref{SxSx}),
where $|\psi(g)\ra$ and $|\psi(h)\ra$ are both matrix product states
with matrices $A_i(g)$ and $A_i(h)$ respectively.  For the operator
in question, namely $S_{x}$, we find from
$$S_x=\frac{1}{\sqrt{2}}(|0\ra\la 1|+|0\ra\la \overline{1}|+|1\ra\la
0|+|\overline{1}\ra\la 0|)$$ the form of its matrix operator,
\begin{equation}\label{}
    \tilde{E}_{S_x}=\frac{1}{\sqrt{2}}(gI\otimes (A_1+A_{\overline{1}})+(A_1+A_{\overline{1}})\otimes
    hI),
\end{equation}
which can be simply written as
\begin{equation}\label{}
    \tilde{E}_{S_x}=\frac{1}{\sqrt{2}}(gX_2+hX_1),
\end{equation}
where $X=A_1+A_{\overline{1}}$ and the indices on $X$ means its
embedding on the first and second spaces. Thus we obtain
\begin{equation}\label{}
\la
\psi(g)|S_{x,1}S_{x,r}|\psi(h)\ra=\frac{1}{2}tr((gX_2+hX_1)(V+ghI)^{r-2}(gX_2+hX_1)
(V+gh)^{N-r-2}).
\end{equation}
Expanding the right hand side and collecting the coefficient of
$(gh)^n$ will give
\begin{equation}\label{x1xrn}
\frac{\la \psi_n|S_{x,1}S_{x,r}|\psi_n\ra}{\la \psi_n|\psi_n\ra} =\frac{1}{2}\frac{\sum_{k=0}^{r-2}\left(\begin{array}{c}r-2\\
k\end{array}\right)\left(\begin{array}{c}N-r-2\\
n-1-k\end{array}\right)tr(\Omega^x_{N,r,k,n})}{\left(\begin{array}{c}N\\
n\end{array}\right)tr(V^{N-n})},
\end{equation}
where
\begin{equation}\label{Omega}
    \Omega^x_{N,r,k,n} := X_2V^{r-2-k}X_1V^{N-r-n-1+k}+X_1V^{r-2-k}X_2V^{N-r-n-1+k}
\end{equation}

We now come to the thermodynamic limit of these correlation
functions.

\subsubsection{The thermodynamic limit} In this subsection we
calculate the thermodynamic limit of the correlation functions found
above, by which we mean taking $N\lo \infty$ while keeping $r$ and
$n$ fixed. In this limit, the sums in the numerators of equations
(\ref{z1zrn}) or (\ref{x1xrn}) are dominated by their $k=0$ term,
therefore we find from (\ref{z1zrn})
\begin{equation}\label{}
    \frac{\la \psi_n|S_{z,1}S_{z,r}|\psi_n\ra}{\la
    \psi_n|\psi_n\ra}=\frac{tr(UV^{r-2}UV^{N-r-n})}{tr(V^{N-n})}.
\end{equation}
Moreover when taking the traces of $V^N$, we need only keep the
eigenvalue with largest magnitude. The matrix $V$ has two
eigenvalues with largest magnitudes, namely $\lambda_{\pm}=\pm
\sqrt{2}$, whose corresponding eigenvectors we denote by $|\pm\ra$.
Their explicit form is
\begin{equation}\label{pm}
  |\pm\ra = \frac{1}{2}(|11\ra\pm
  \sqrt{2}|00\ra+|\overline{1}\overline{1}\ra),\h \lambda_{\pm}=\pm
  \sqrt{2}.
  \end{equation}

 Therefore by using the fact that $N$ and $n$ are even
 (see the discussion preceding equation (\ref{psigggg})),
 we arrive at
\begin{equation}\label{}
    \frac{\la \psi_n|S_{z,1}S_{z,r}|\psi_n\ra}{\la
    \psi_n|\psi_n\ra}=\frac{1}{2}\frac{\la + |UV^{r-2}U|+\ra + (-1)^r\la -| UV^{r-2}U|-\ra}{2^{r/2}}.
\end{equation}
Using equations (\ref{mpsop}) and (\ref{pm}) and noting the explicit
form of the matrices $A_0=|1\ra\la 0|+|0\ra\la \overline{1}|$ and
$A_1=|0\ra\la 1|+|\overline{1}\ra\la 0|$, we find after a
straightforward calculation, that  the above matrix elements vanish
unless $r=2$. Using the fact that $\la \pm |U^2|\pm\ra=-1$, this
will give the correlation function
\begin{equation}\label{}
    \frac{\la \psi_n|S_{z,1}S_{z,r}|\psi_n\ra}{\la
    \psi_n|\psi_n\ra}=-\frac{1}{2}\delta_{r,2}.
    \end{equation}
Similarly for the correlation of transverse components we find from
(\ref{x1xrn}) and following a similar reasoning as above that
\begin{equation}\label{}
    \frac{\la \psi_n|S_{x,1}S_{x,r}|\psi_n\ra}{\la
    \psi_n|\psi_n\ra}=\frac{1}{2}\frac{n}{N}\frac{tr(\Omega^x_{N,r,0,n})}{tr(V^{N-n})},
\end{equation}
which simplifies to
\begin{equation}\label{}
    \frac{\la \psi_n|S_{x,1}S_{x,r}|\psi_n\ra}{\la
    \psi_n|\psi_n\ra}=\frac{n}{N}\frac{1}{2^{\frac{r+1}{2}}}
    (\la + |\Omega^x|+\ra + (-1)^{r-1}\la -|\Omega^x|-\ra),
\end{equation}
where $\Omega^x=X_1V^{r-2}X_2+X_2V^{r-2}X_1$. Obviously this tends
to zero in the limit we are considering, due to the pre-factor
$\frac{n}{N}$.

\section{Discussion}
We have made a detailed study of two new spin systems whose ground
states can be found exactly within the matrix product formalism, and
have shown that the method of MPS formalism may be more fruitful
than it appears at first sight.  In the space of quantum spin chains
solvable by this formalism, there may be physically interesting
models with rich properties. For example, there may be ground states
which break the continuous symmetries of their parent Hamiltonian,
or one may find matrix product states which capture as a generating
state, a large number of degenerate ground states of a given
Hamiltonian. In this paper we have made an exhaustive study of
spin-1 matrix product states with three-dimensional auxiliary
matrices and have found two distinct class of models. The first
model is a one-parameter family of models interpolating between the
Ising-like Hamiltonnian
$$H_1=\sum_{i}(S^2_{z,i}-1)(S^2_{z,i+1}-1), $$ and the
rotationally invariant Hamiltonian
$$H_2=\sum_{i}({\bf S}_i\cdot{\bf
S}_{i+1})^2.$$ The ground state of this latter Hamiltonian which is
in the form of an MPS, breaks the rotational symmetry SO(3) of $H_2$
to that of SO(2). Symmetry generators now will give other degenerate
ground states of $H_2$. The second model which we have found
corresponds to a kind of $6$-vertex model, already studied in
\cite{KlumperRep1}. Here the MPS is a generating state of the
degenerate ground states of the Hamiltonian, none of which may have
a simple MPS representation. By suitable manipulations of this
generating state, we have been able to find one and two-point
functions for these ground states, which are other-wise very
difficult to calculate.

\section{Acknowledgements}
We would like to thank the partial financial support of center of
excellence in Complex Systems and Condensed Matter (CSCM) for this
project.

\section{Appendix A}
In this appendix we show explicitly for model II, that the two
different choices $h=g$ and $h=-g$ lead to equivalent models. A
similar analysis applies to model I, which we omit here. Let
$h=\sigma g$, where $\sigma=\pm 1$. Then it can be verified that the
null space ${\cal V}_2$ is spanned by the vectors:
\begin{eqnarray}\label{e}
|e_1\ra=\frac{1}{\sqrt{2}}(|01\ra-\sigma|10\ra),\cr
|e_2\ra=\frac{1}{\sqrt{2}}(|0\overline{1}\ra-\sigma|\overline{1}0\ra).
\end{eqnarray}
Writing the local Hamiltonian as usual $h_{II}=|e_1\ra\la
e_1|+|e_2\ra\la e_2|$, we see that
\begin{equation}\label{rot}
    h_{II}(\sigma=-1)=(I\otimes R_{z}(\pi))h_{II}(\sigma=1)(I\otimes
    R_{z}(\pi)^{-1}),
\end{equation}
where $R_{z}(\pi)$ is a rotation around $z$ axis by $\pi$ degrees.
Thus we arrive at
\begin{equation}\label{}
    H_{II}(-\sigma)=\mathcal{R} H_{II}(\sigma) \mathcal {R}^{-1},
\end{equation}
where we have taken $N$ to be an even number and $\mathcal{R}$ is a
global rotation of the type
\begin{equation} \mathcal{R}=\bigotimes_{i=0} R^{2i+1}_{z}(\pi).
\end{equation}
This shows that $H_{II}(\sigma)$ and $H_{II}(-\sigma)$ are
iso-spectral and have the same thermodynamic properties and for this
reason we have considered only the model $H_{II}(\sigma=1)$.
{}
\end{document}